\documentclass[10pt,letterpaper]{article}
\usepackage[top=0.85in,left=2.75in,footskip=0.75in]{geometry}

\usepackage{changepage}

\usepackage[utf8]{inputenc}

\usepackage{textcomp,marvosym}

\usepackage{fixltx2e}

\usepackage{amsmath,amssymb}

\usepackage{cite}

\usepackage{nameref,hyperref}

\usepackage[right]{lineno}

\usepackage{microtype}
\DisableLigatures[f]{encoding = *, family = * }

\usepackage{rotating}

\usepackage[aboveskip=1pt,labelfont=bf,labelsep=period,singlelinecheck=off]{caption}

\makeatletter
\renewcommand{\@biblabel}[1]{\quad#1.}
\makeatother

\date{}

\usepackage{lastpage,fancyhdr,graphicx}
\usepackage{epstopdf}
\fancyheadoffset[L]{2.25in}
\fancyfootoffset[L]{2.25in}

\begin{document}
\vspace*{0.35in}

\begin{flushleft}
{\Large
\textbf\newline{The Impact of Heterogeneous Thresholds on Social Contagion with Multiple Initiators}
}
\newline
\\
Panagiotis D. Karampourniotis\textsuperscript{1,2*},
Sameet Sreenivasan\textsuperscript{1,2},
Boleslaw K. Szymanski\textsuperscript{2,3,4},
Gyorgy Korniss\textsuperscript{1,2}
\\
\bigskip
\bf{1}  Department of Physics, Applied Physics, and Astronomy, Rensselaer Polytechnic Institute,
110 8$^{th}$ Street, Troy, NY, 12180-3590 USA
\\
\bf{2} Social Cognitive Networks Academic Research Center,
Rensselaer Polytechnic Institute, 110 8$^{th}$ Street, Troy, NY, 12180-3590 USA
\\
\bf{3} Department of Computer Science,
Rensselaer Polytechnic Institute, 110 8$^{th}$ Street, Troy, NY, 12180-3590 USA
\\
\bf{4} Spo\l{}eczna Akademia Nauk, \L{}\'{o}d\'{z}, Poland
\\
\bigskip
* karamp@rpi.edu

\end{flushleft}
\section*{Abstract}
The threshold model is a simple but classic model of contagion
spreading in complex social systems. To capture the complex nature
of social influencing we investigate numerically and analytically
the transition in the behavior of threshold-limited cascades  in the
presence of multiple initiators as the distribution of thresholds is
varied between the two extreme cases of identical thresholds and a
uniform distribution. We accomplish this by employing a truncated
normal distribution of the nodes' thresholds and observe a
non-monotonic change in the cascade size as we vary the standard
deviation. Further, for a sufficiently large spread in the threshold
distribution, the tipping-point behavior of the social influencing
process disappears and is replaced by a smooth crossover governed by
the size of initiator set. We demonstrate that for a given size of
the initiator set, there is a specific variance of the threshold
distribution for which an opinion spreads optimally.
Furthermore, in the case of synthetic graphs we show that the spread asymptotically
becomes independent of the system size, and that global cascades
can arise just by the addition of a single node to the
initiator set.



\section*{Introduction}
The technological breakthroughs of the 21st century have strongly
contributed to the emergence of network science, a multidisciplinary
science with applications in many scientific fields and
technologies. Several sociological opinion diffusion models first
introduced in the middle of 20th century are now being thoroughly
studied, while variations of these classical models have been
introduced. Most of these models are based on social reinforcement,
where simple rules based on the interaction of individuals with
their respective nearest neighbors govern individual opinion
evolution. The macroscopic outcome of these rules is a cascade of
nodes switching opinions~\cite{Schelling,Granovetter,Latane_PR1990,Latane_1996,Watts_2002,Watts_2007,Centola_2007,Xie,Ugander}.
We focus our study on one of the classic models of social
influencing, the Threshold Model (TM). The TM is a binary opinion
spread model first introduced by Granovetter~\cite{Granovetter} to
model collective behavior socially driven by peer pressure. Under the TM a node adopts a new opinion only when the
fraction of its nearest neighbors possessing that opinion is larger
than an assigned threshold, which represents the resistance of the
node to peer pressure~\cite{Latane_PR1990}.
Although the microscopic rule of opinion adoption in the TM is
simple, the collective behavior that arises is complex and
non-linear. The resulting spread size depends on a large set of
parameters, such as the network structure (e.g.,
clustering)~\cite{Centola_2007,Singh,Ikeda_2010,Lee_2014,Nematzadeh_2014},
the size of the initially active nodes (initiators), the selection
strategy of the initiators, and the distribution of threshold values among nodes of the network. The first thorough investigation of the TM was made by
Watts~\cite{Watts_2002}, who examined the effect of one randomly
selected initiator on the cascade size. Gleeson and
Cahalane~\cite{Gleeson_2007,Gleeson_sigma,Gleeson_time}, on the other hand,
determined analytically the cascade size for varying initiator sizes
(or fractions) for the infinite system size. Recent investigations of
the TM by Karimi and Holme~\cite{Holme_PhysA2013} and Michalski et al.~\cite{Michalski_NGC2014} also considered
the impact of temporal networks on contagion cascades.
Very recently, Ruan et al.~\cite{Ruan_2015} studied the effects of ``immune" individuals (those who resist adopting the new idea indefinitely) and external influencing (e.g., by mass media or advertisements) in the TM.

An important problem in generalized models for social and biological
contagion~\cite{Dodds_JTB2005,Kitsak_NatPhys2010,Karsai_JRSI2014} is
to optimize the set of initiators, i.e., for a fixed cost (seed
size), find the set of initiators giving rise to the largest
cascade, or alternatively, find the minimum size seed set required
to activate the entire network~\cite{Flaviano_2015}. As far as
selection strategies are concerned, Kempe et al.~\cite{Kempe_2003}
showed that the optimization problem of selecting the most
influential nodes in any directed weighted graph with uniform random
selection of thresholds is NP-hard. They also suggested a greedy
algorithm~\cite{Kempe_2003}, where each new initiator is selected
based on the maximum spread it can cause, which unfortunately
resulted in low efficiency of the algorithm. Chen et
al.~\cite{Chen_2010} designed a scalable algorithm (LDAG) which is
based on the properties of directed acyclic graphs. Recently, Lim et
al.~\cite{Lim_2015} introduced a new node-level measure of
influence, called cascade centrality (based on the size of the
cascade resulting from the node being the only initiator), which may
guide the selection of multiple initiators. Closely related to
these studies and of practical interest is to find a set of initiators
(not necessarily the smallest) in a scalable fashion that guarantees
that the entire network will ultimately turn active, triggered by
these initiators~\cite{Shakarian_SNAM}. Their method was inspired by
the $k$-shell decomposition of the network~\cite{Carmi_PNAS2007},
which itself can be an effective heuristic for selecting
initiators in a broad class of models for the spreading of social or
biological contagion~\cite{Kitsak_NatPhys2010}.

Singh et al.~\cite{Singh} studied the effect in the TM of varying the
fraction of initiators on the cascade size for various basic
heuristic selection strategies when each node has identical
threshold in the network. They showed that there is a critical
fraction of initiators (``tipping point") at which a sharp
(discontinuous) phase transition occurs from small to large cascades
in Erd\H{o}s-R\'enyi (ER) graphs~\cite{Erdos_1959}. This phase transition is apparent
for the random, $k$-shell, and degree-ranked selection strategies,
which are listed in the increasing order of their performance. These findings, in
particular, the emergence of the discontinuous transition, were
analogous to those found by Baxter et al.~\cite{Baxter_PRE2010} for
bootstrap percolation (there, activation of a node requires $k$
active neighbors).

Watts~\cite{Watts_2002}, proposed the first analytic solution for
the TM, using percolation theory and generating functions to measure
the size of the largest cluster of nodes requiring only one active
neighbor to turn active (largest vulnerable cluster). The model
applies to unweighted, undirected graphs with small clustering
coefficient. In the infinite system size, when the vulnerable
cluster percolates, there is a non-zero probability that a cascade
will take over a large portion of the network (global cascade). A
randomly selected initiator will activate the largest vulnerable
cluster, if it is a part of the cluster or is one of its neighbors. Using this
analytic method, Watts studied the regime
for which global cascades are possible for one initiator, for different values of identical
thresholds $\phi_0$ and average degree $z$ of synthetic graphs. He found that, for ER graphs with {\it O}(1)
initiator the criterion for global cascades is $z<1/\phi_{0}$.

Gleeson and Cahalane~\cite{Gleeson_2007} formulated an analytic
approach for the TM with varying initiator sizes. Their work was
inspired by the zero-temperature Random-Field Ising Model (RFIM)~\cite{Dhar,Sethna}, where the cascade size, the initiator size and
the threshold distribution correspond to the magnetization, the
external uniform field and the local quenched random fields of the
RFIM. The main difference between the two models is that in the TM
the activated nodes remain activated, while in the RFIM the spins
may flip back to an inactive state. The analytic approach to the TM model is
applicable to locally tree-like structures~\cite{Gleeson_2007}, such as ER graphs. The
graph is considered an infinite-level tree with a level-by-level
updating of the spread size, starting from the bottom of the tree.

In most of the past research, the cascade size has been thoroughly
investigated for a identical threshold in the
network~\cite{Watts_2002,Ikeda_2010, Lee_2014,
Nematzadeh_2014,Singh}, or for a random threshold for each
node~\cite{Kempe_2003,Chen_2010}. However, a model with identical
thresholds does not capture the complex nature of social influencing
when multiple initiators are present. The small scale experiment
conducted by Latane~\cite{Latane_1996} and more recently an online
experiment by Centola~\cite{Centola_2010} and a large online study on Facebook data ~\cite{State_2015}
suggest that individuals have diverse thresholds for adopting a newly introduced opinion.
Here, to capture the diversity of opinion adoption thresholds in a
social influence context, we study the effect of heterogeneous
thresholds on the cascade size under the TM for empirical and synthetic
unweighted and undirected networks for randomly selected initiators.

\section*{Materials and Methods}

\subsection*{Simulations of the Threshold Model}
We assume that the thresholds are drawn from a truncated
normal distribution with mean $\phi_{0}$ and standard deviation
$\sigma$. The threshold $\phi$ of each node is limited to interval [0,1],
thus the mean threshold  $\phi_{0}$ is
also within this interval, and $\sigma$ is in the range of [0,
0.288], boundaries of which correspond to the identical threshold
and to the random threshold, respectively.
Unlike, in the formulation of the threshold model
in~\cite{Gleeson_2007,Gleeson_sigma}, where thresholds drawn can be
negative, allowing nodes to get spontaneously activated as
innovators, and as a result randomizing the set of initiators, we
are interested in the case where spread is initiated only with
the insertion of randomly selected initiators in the network.

Once a threshold for each node is set, for the simulations, we
randomly assign initiators one by one and measure the cascade size.
We repeat this process by drawing thresholds
from the same distribution. The final cascade size for each
threshold distribution is obtained by averaging one thousand times
on different threshold distribution draws and, for the synthetic
graphs, different network realizations.

\subsection*{Network Structures}
The networks we use are undirected and unweighted. The synthetic networks used are Erd\H{o}s-R\'enyi (ER) graphs
and scale-free (SF) networks.
For the generation of ER graphs~\cite{Erdos_1959} we used the $G(N,p_{\rm ER})$ model with $N$ being the system size and $p_{ER}$ the probability that a random node will be connected to any node in the graph. The probability $p_{ER}$ is given by $p_{ER}=z/\left(N-1\right)$, where $z$ is the nominal average degree in the network. We keep the average degree $z=10$.
For the generation of uncorrelated SF networks~\cite{Barabasi_1999,Catanzaro} ($N=10^4$, $z=10$, with power law constant $\gamma=3$) we employ the configuration model~\cite{Britton,Catanzaro} with a structural cut-off,  and a maximum possible node degree set to $\sqrt{N}$, using a high accuracy look-up table from~\cite{Molnar_MDS}.

The empirical networks used are a connected ego-network from a Facebook (FB) dataset, available from the Stanford Network Analysis Project (SNAP)~\cite{SNAP} (system size $N=4048$, average degree $z=43$), and
a high-school (HS) friendship network~\cite{Add_Health}.
For the HS network, we only used the giant connected component of that network, with $N=921$ and $z=5.96$. The network contains two communities which are roughly equal in size (for more information on the two empirical networks see table in S1 Text).
Although SF, FB, and HS networks are connected networks, the generated ER
graphs may have a disconnected component with probably $e^{-z}$,
which for $z=10$ is approximately 0.000045.

\subsection*{Tree-like approximation for the Threshold Model}
For analytic methods, we apply Gleeson's and Cahalane's
tree-like approximation for synthetic networks~\cite{Gleeson_2007,Gleeson_sigma}.
The approximation is given by the following set of equations
\begin{equation}
S_{eq}=p+\left(1-p\right)\sum\limits_{k=1}^\infty P_k\sum\limits_{m=1}^k \binom {k} {m}q_\infty^m\left(1-q_\infty\right)^{k-m}F\left(\frac{m}{k}\right)
\label{final_spread_TL}
\end{equation}
\begin{equation}
q_{n+1} = p+\left(1-p\right)\sum\limits_{k=1}^\infty \frac{k}{z}P_k\sum\limits_{m=1}^{k-1} \binom {k-1} {m}q_n^m\left(1-q_n\right)^{k-m-1}F\left(\frac{m}{k}\right).
\label{n_level_spread_TL}
\end{equation}
In this approximation the graph is considered an infinite level
tree. The spread diffuses level-by-level starting from the bottom of
the tree. $q_n$ is defined ``as the conditional probability that a node on
level $n$ is active, conditioned on its parent on level $n+1$
being inactive" and it is given by Eq.~(\ref{n_level_spread_TL}). The final spread $S_{eq}$ is given by
Eq.~(\ref{final_spread_TL}), and is measured at the top of the tree.
The fraction of initially active nodes is given by $p$.
In the bottom of the tree at level $n=0$, the fraction of active nodes is only
based on the initiators, thus $q_0=p$.
The graph degree distribution is given by $P_k$, which for an
infinite size ER graph is given by $P_k=\left(z^k e^{-z}\right)/k!$, where $z$
is the average degree, while for SF networks it's given by $P_k\sim k^{-\gamma}$. $F\left(\frac{m}{k}\right)$ is the cumulative
probability that a node requires m or less active neighbors to get
active, which depends on the assigned threshold distribution.

\section*{Results}
\begin{figure}[h]
\centerline{\includegraphics[width=\textwidth]{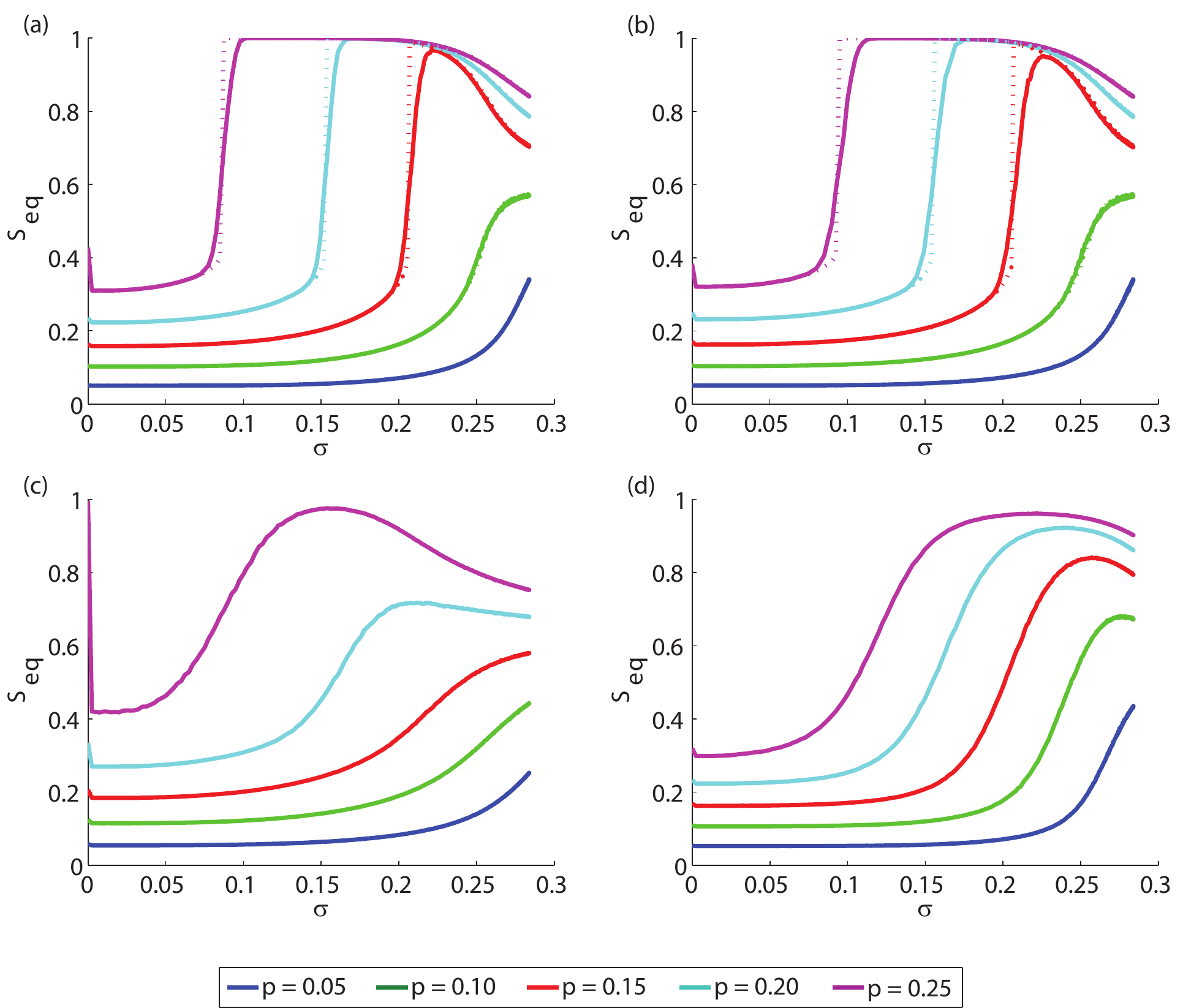}}
\vspace{0.3truecm}
\caption{{\bf Behavior of the cascade size $\boldsymbol{S_{eq}}$ at equilibrium for varying standard deviation $\boldsymbol{\sigma}$.}
(a) ER graphs with $z=10$ and $N =10^4$;
(b) SF networks with $z=10$, $\gamma=3$, and $N=10^4$;
(c) high-school network with $z=5.96$ and $N=921$;
(d) facebook network with $z=43$ and $N=4039$.
The mean threshold is $\phi_0=0.50$.
The simulations are averaged over one thousand repetitions.
(a) and (b) also show the analytic estimates (dotted lines) based on the tree-like approximation (see Materials and Methods) \cite{Gleeson_2007}.}
\label{Fig1}
\end{figure}

\begin{figure}[h]
\centerline{\includegraphics[width=\textwidth]{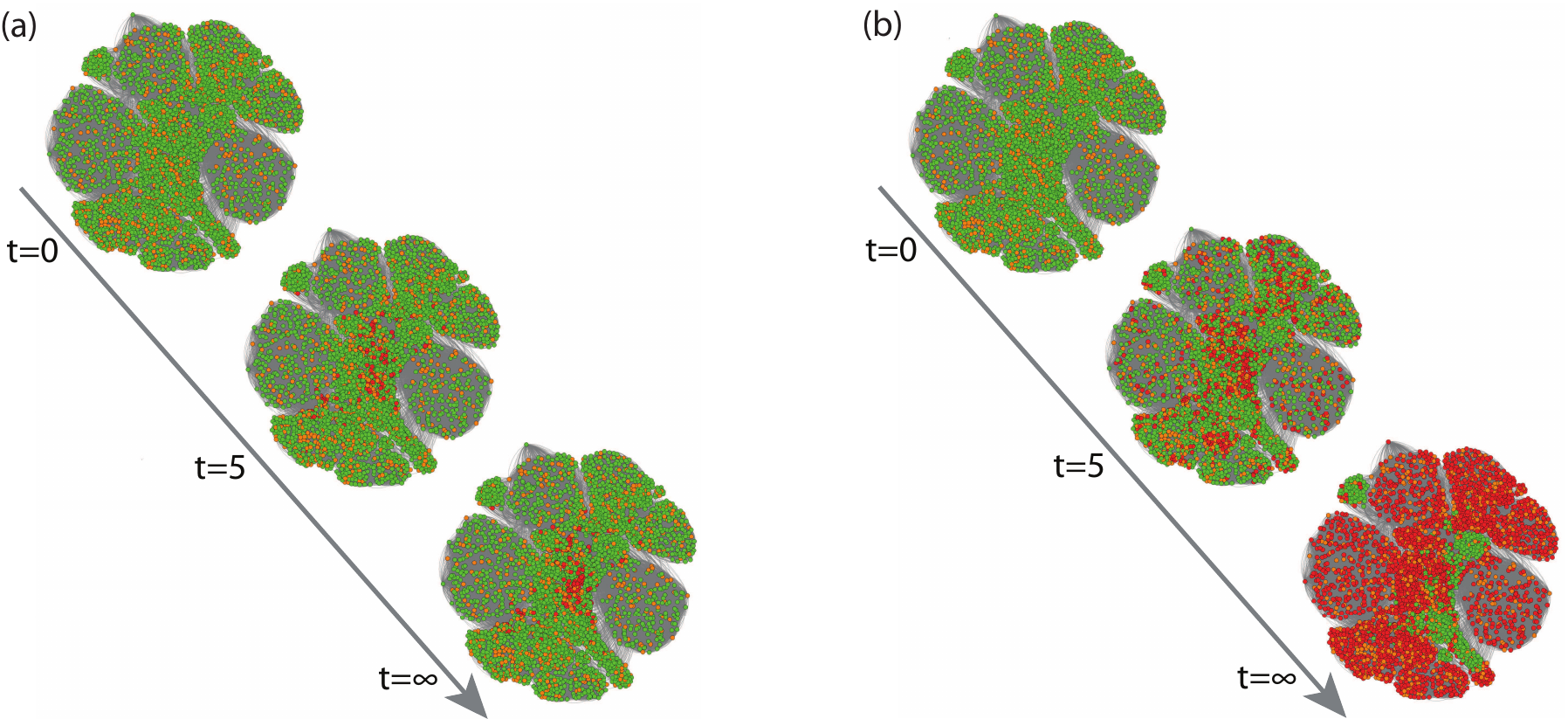}}
\vspace{0.5truecm}
\caption{{\bf Visualization of the spread of opinion in the TM model on a facebook network with $\boldsymbol{z}$$\boldsymbol{=}$$\boldsymbol{43}$ and $\boldsymbol{N}$$\boldsymbol{=}$$\boldsymbol{4039}$.}
The fraction of the randomly selected initiators is $p=0.20$.
The mean threshold is $\phi_0=0.50$ while the standard deviation of the threshold is (a) $\sigma=0$, (b) $\sigma=0.20$.
Inactive nodes, initiators, and active nodes (through spreading) are marked with green, orange, and red, respectively.}
\label{Fig2}
\end{figure}

First, we examine the effect of the standard deviation $\sigma$ on
the cascade size $S_{eq}$ (averaged) for a constant initiator fraction and
constant mean threshold $\phi_0$ (Fig.~\ref{Fig1}). As $\sigma$ increases so does a fraction of nodes whose threshold is far from the average causing a twofold effect.
Of nodes far from average, the ones with thresholds
below average are easily activated while those with thresholds above
average are increasingly difficult to activate. Thus, when the
initiator fraction is small, the cascade size $S_{eq}$ is
monotonically increasing since the presence of larger fraction of low
threshold nodes facilitate the spread. However, when the initiator
fraction are large, the increase in low threshold nodes helps a
little since they are likely to be already activated without the
increase in $\sigma$, but presence of additional high threshold
nodes arrest the spread. This trade-off gives rise to the
non-monotonic behavior seen in Fig.~\ref{Fig1}, which is apparent
for different types of networks. Depending on the network structure
and size of the initiators, the standard deviation $\sigma$ for
which the spread is optimal varies.
A visualization (Fig.~\ref{Fig2}) shows time steps of the spread
on a random selection of initiators with $p=0.20$ in the FB network.
For the same set of initiators, the spread for large sigma ($\sigma=0.20$)
is much higher than for identical thresholds ($\sigma=0.00$).
Interestingly, in the vicinity of $\sigma \sim 0$ the sharp decrease in the cascade
size $S_{eq}$ occurs because with non-zero $\sigma$, approximately
half of the nodes acquire a threshold higher than $\phi_0=0.50$. For all the nodes with threshold $\phi>\phi_0$ with even
degree, even the slightest non-zero $\sigma$ value will increase the number of active neighbors  by one, thus making cascades less likely to occur.
Finally, for ER graphs [Fig.~\ref{Fig1}(a)] and SF networks [Fig.~\ref{Fig1}(b)] the analytic estimates are in good agreement with the simulations.

\begin{figure}[h]
\centerline{\includegraphics[width=\textwidth]{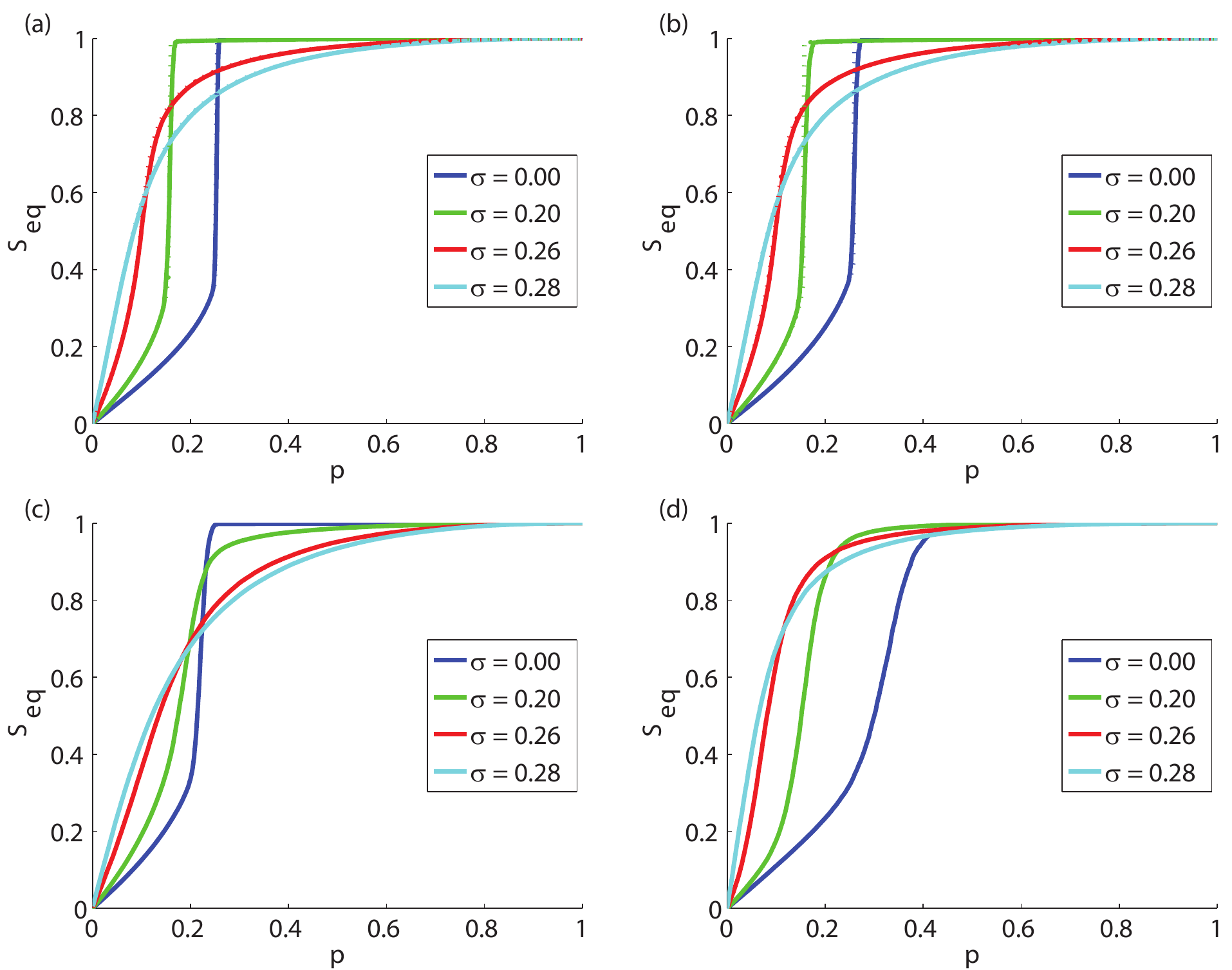}}
\vspace{0.3truecm}
\caption{{\bf Behavior of the cascade size $\boldsymbol{S_{eq}}$ at equilibrium vs. the initiator fraction $\boldsymbol{p}$.}
The networks are the same as in Fig.~\ref{Fig1}:
(a) ER graphs with $z=10$ and $N =10^4$;
(b) SF networks with $z=10$, $\gamma=3$, and $N=10^4$;
(c) high-school network with $z=5.96$ and $N=921$;
(d) facebook network with $z=43$ and $N=4039$.
The mean threshold is $\phi_0=0.50$.
(a) and (b) also shows the analytic estimates (dotted lines) based on the tree-like approximation (see Materials and Methods)\cite{Gleeson_2007}.}
\label{Fig3}
\end{figure}

In Fig~\ref{Fig3}, the cascade size $S_{eq}$ is plotted for
varying initiator sizes $p$ for the same networks as in
Fig.~\ref{Fig1}. As the initiator fraction increases, for small
enough $\sigma$ there is a transition from small local cascades to
large global cascades, which, for synthetic graphs is a
discontinuous phase transition [Fig.~\ref{Fig3} (a) and (b)].
However, the line of the average cascade size $S_{eq}$ appears
smooth even in the presence of a discontinuous phase transition,
because for each repetition the point of the discontinuous phase
transition varies slightly. With increasing $\sigma$ the initiator
fraction for which the transition occurs is reduced, while for the
synthetic graphs the spread size still exhibits a discontinuous
phase transition. With largely diverse thresholds we find that a
critical initiator size beyond which cascades become global ceases
to exist and the tipping-point behavior of the social influencing
process disappears and is replaced by a smooth crossover governed by
the size of initiator set. This property can be important, for example,
for a company's marketing strategy of a new product. If the
threshold distribution is narrow enough, unless a critical initiator
fraction is reached, there is a marginal local spread on a few of
the first or second neighbor friends of the initiators. On the other
hand, if the threshold distribution is wide, there is a significant
spread. For the uniform random threshold distribution each addition
of initiators has a reduced contribution to the cascade size as
predicted by the submodularity property of the TM~\cite{Kempe_2003}.

 In Figs~\ref{Fig4} and \ref{Fig5} we show
that the behavior of the cascade size is largely independent of the system size $N$
for any threshold distribution with the same  degree distribution, for ER graphs and
SF networks, respectively. We observe that with increasing system size $N$
the cascade size $S_{eq}$ is asymptotically  converging.

We record the critical initiator fraction $p_c$ for which a discontinuous phase transition occurs for varying mean threshold $\phi_0$ (Fig.~\ref{Fig6}). For the measurement of $p_c$, first we calculated the derivative of the $S_{eq}$ from Fig.~\ref{Fig3} with respect to the initiator fraction $p$. The position of maximum of the derivative yields the $p_c$, in other words, $p_c=\arg\max_p\left(dS_{eq}\left(p\right)/dp\right)$. We used the same method for the calculation of the respective analytic estimates.
We confine the threshold distribution for up to $\sigma=0.15$ to assess if there is a discontinuous phase transition with increasing initiators.
Above each $p_c$ line global cascades occur. The value of $p_c$ decreases with increasing $\sigma$.
For identical thresholds $\phi_0$ (in blue), the $p_c$ line has some sharp jumps, for example at $\phi_0$ equal to 0.50, 0.33, and 0.25 (Fig.~\ref{Fig6}).
These jumps are artifacts of the discrete steps of the degree distribution in the presence of a unique threshold for all the nodes. In particular, microscopically, the number of active neighbors required for a node to turn active
increases by integer values. For example, for a node with degree 10
and $0.40<\phi\leq0.50$, that number is 5. For identical thresholds in the
network, the cumulative effect of these integer steps gives rise to the
jumps exhibited by the $p_c(\phi_{0})$ curves (Fig.~\ref{Fig6}).
Interestingly, this effect also shows in Fig.~\ref{Fig1}, where
for large enough initiator fractions (i.e., $p=0.25$ or higher) the cascade size drops abruptly as $\sigma$ is increased from zero to small values.
For nodes with mean threshold
$\phi_0=0.50$, even the  smallest non zero increase on the standard
deviation $\sigma$ results in approximately half of the nodes
having threshold larger than $\phi_0=0.50$. The $p_c$ lines are lower
for the ER graph compared to the SF networks because of the
importance of a randomly selected very high degree node in SF
networks can have on the spread. Our results obtained
from simulations are in agreement with the analytic
estimates.

To further understand the effect of the standard deviation $\sigma$,
we study the dynamics of the spread for synchronous updating of the nodes.
In phase-space, as shown in Fig.~\ref{Fig7}, the difference
$\Delta S(n+1)-\Delta S(n)$ defines the number of nodes activated from time step $n$ to $n+1$.
The dynamic spread in the TM is deterministic and evolves in one direction, hence, the spread stops when the change on the cascade size ($Y$-axis) reaches zero. Accordingly, the value of the cascade size in the steady state is indicated on the
$X$-axis. When cascades are not possible, the spread rate decreases
monotonically. However, when cascades are possible then for up to
some $\sigma$ the change is non-monotonic and the fractions of nodes
in cascades reach almost one. But as $\sigma$'s grow larger and
larger, these fractions stop growing farther and stay farther from one.
When $\sigma$ approaches the standard deviation of uniform
distribution the shape of the lines decreases linearly.
Interestingly, similar behavior is observed for the FB and HS
networks as well.

\subsection*{Closed-form analytic estimate for the uniform threshold distribution}
For a uniform threshold distribution the phase-space line decreases
linearly for any initiator fraction for synthetic graphs and almost
linearly for the empirical networks (Fig.~\ref{Fig8}). In addition, we show for this threshold distribution,
using Gleeson's and Cahalane's analytical methods, that the phase-space line has a closed form and is
linearly decreasing. The extended proof of this is shown in S1 Text. 
For a uniform threshold distribution the iterative formula in Eq.~(\ref{n_level_spread_TL})
of the analytic approximation yields the following closed-form solution
\begin{equation}
q_{n+1} = p+bq_n,
\label{q_n linear}
\end{equation}
with $b=\left(1-p\right)\frac{1}{z}\left(z-1+ P_0\right)$. The solution of the above iterative equation with the initial condition $q_0=p$, is
\begin{equation}
q_n = p\frac{1-b^{n+1}}{1-b}.
\label{q_n solution}
\end{equation}
According to \cite{Gleeson_time},
the spread at level $n+1$ is given by
\begin{equation}
S_{n+1}=h(q_n)=p+\left(1-p\right)\sum\limits_{k=1}^\infty P_k\sum\limits_{m=1}^k \binom {k} {m}q_n^m\left(1-q_n\right)^{k-m}F\left(\frac{m}{k}\right),
\label{S_{n+1} formula}
\end{equation}
which, in the case of a uniform distribution of thresholds (S1 Text) simplifies to
\begin{equation}
S_{n+1} = p + cq_n,
\label{S_{n+1} vs q_n}
\end{equation}
with $c=\left(1-p\right)\left(1-P_0\right)$, where the initial spread is $S_0=p$.
Using the above Eq. and Eq.~(\ref{q_n linear}) we can calculate (S1 Text) the formula for the phase-space diagram
\begin{equation}
S_{n+1}-S_n =cp-(1-b) p-(1-b)S_n
\end{equation}
The above Eq. is the closed form phase-space line of Fig.~\ref{Fig8}.
On the other hand, at the equilibrium (as $n\rightarrow \infty$) the spread size in Eq.~\ref{S_{n+1} vs q_n} becomes
\begin{equation}
S_{eq} = p + cq_{\infty},
\label{uni_final_cascade_size_TL}
\end{equation}
with $q_{\infty}=p\frac{1}{1-b}$ (S1 Text).
Note that in this approximation for uniform threshold distribution, the size of the final cascade for uncorrelated networks does not dependent on the details of the degree distribution, it only depends on the average degree $z$.
In addition, it is easy to show that the derivative of the final cascade size [Eq.~(\ref{uni_final_cascade_size_TL})] with respect to the initiator size $p$ is monotonically decreasing, in agreement with the submodularity property of the TM for the uniform threshold distribution~\cite{Kempe_2003}.

\subsection*{Discontinuous phase transitions in the threshold model}
\begin{figure}[h]
\centerline{\includegraphics[width=\textwidth]{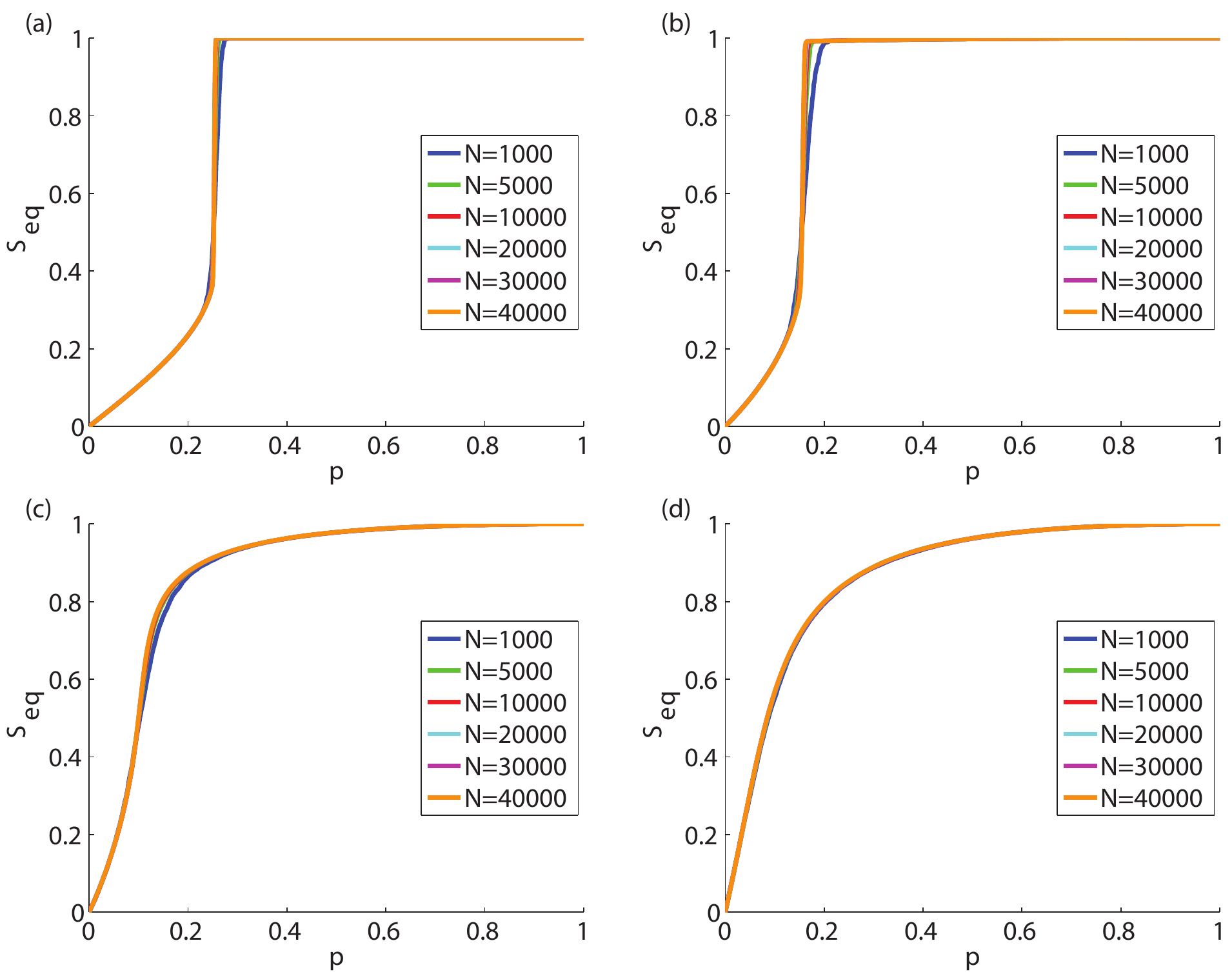}}
\vspace{0.3truecm}
\caption{{\bf Finite-size behavior of the final cascade size $\boldsymbol{S_{eq}}$  vs. the initiator fraction $\boldsymbol{p}$
for ER graphs with average degree $\boldsymbol{z}$$\boldsymbol{=}$$\boldsymbol{10}$.}
The mean threshold is $\phi_0=0.50$ while the standard deviation of the threshold is
(a) $\sigma=0.00$,
(b) $\sigma=0.20$,
(c) $\sigma=0.26$ and
(d) $\sigma=0.28$.}
\label{Fig4}
\end{figure}

\begin{figure}[h]
\centerline{\includegraphics[width=\textwidth]{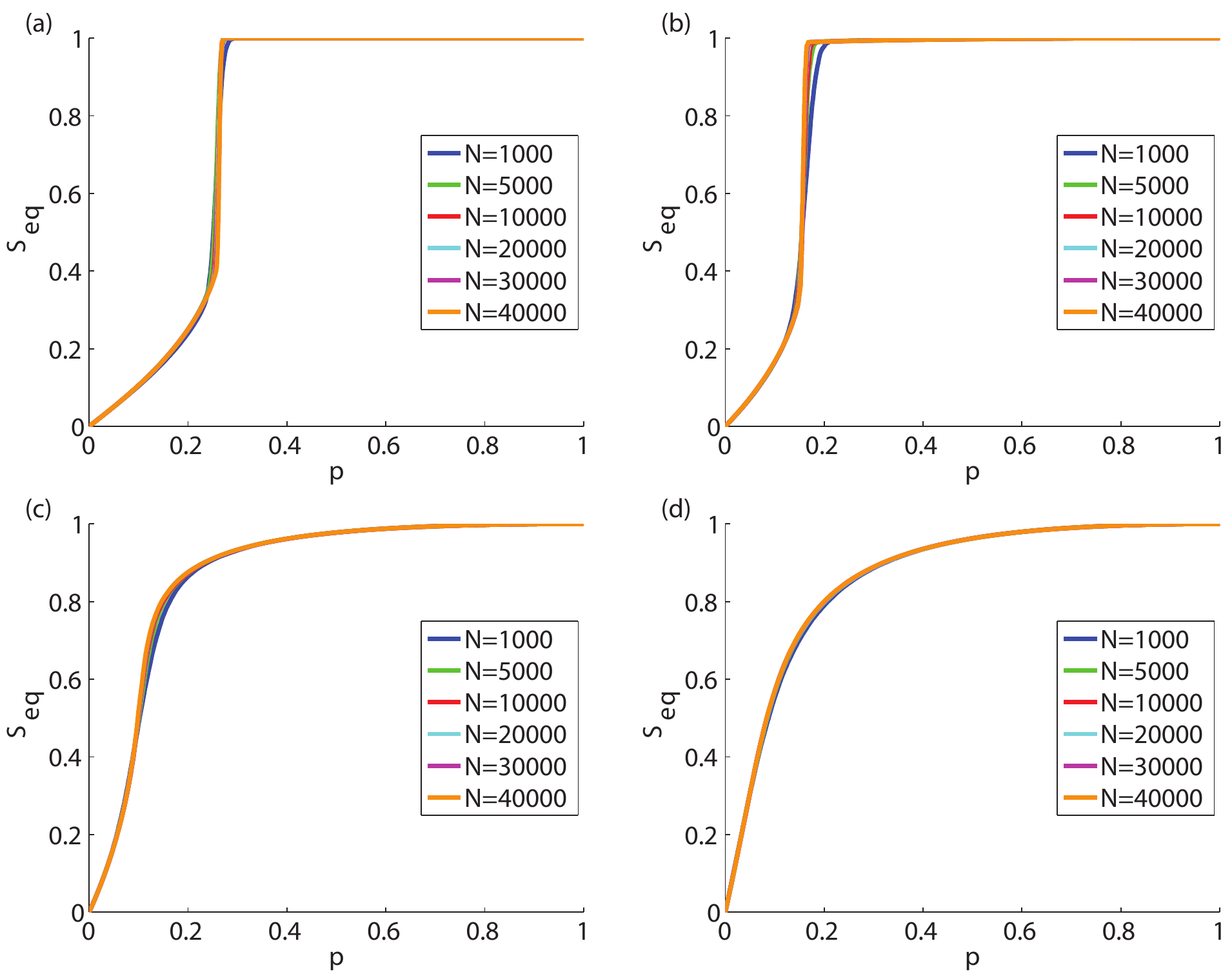}}
\vspace{0.3truecm}
\caption{{\bf Finite-size behavior of the final cascade size $\boldsymbol{S_{eq}}$ at vs. the initiator fraction $\boldsymbol{p}$
for SF networks with $\boldsymbol{z}$$\boldsymbol{=}$$\boldsymbol{10}$ and $\boldsymbol{\gamma}$$\boldsymbol{=}$$\boldsymbol{3}$.}
The mean threshold is $\phi_0=0.50$ while the standard deviation of the threshold is
(a) $\sigma=0.00$,
(b) $\sigma=0.20$,
(c) $\sigma=0.26$,
(d) $\sigma=0.28$.}
\label{Fig5}
\end{figure}

\begin{figure}[h]
\centerline{\includegraphics[width=\textwidth]{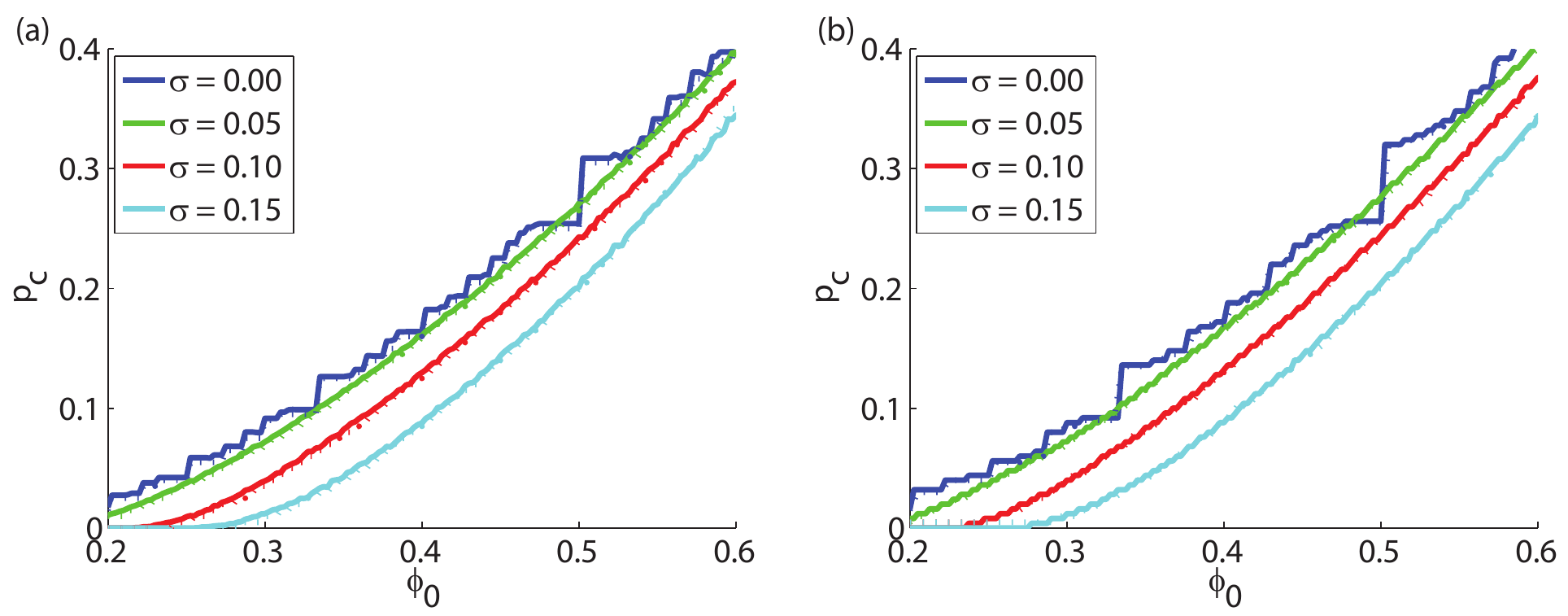}}
\vspace{0.3truecm}
\caption{{\bf Critical initiator fraction $\boldsymbol{p_c}$ vs. mean threshold $\boldsymbol{\phi_0}$.}
(a) ER graphs and (b) SF networks with $\gamma=3$ with average degree $z=10$ and system size $N=10^4$.
An initiator size above the $p_c$ line leads to global cascades.
The analytic estimates (dotted lines) are based on the tree-like approximation \cite{Gleeson_2007} (see Materials and Methods).}
\label{Fig6}
\end{figure}

\begin{figure}[h]
\centerline{\includegraphics[width=\textwidth]{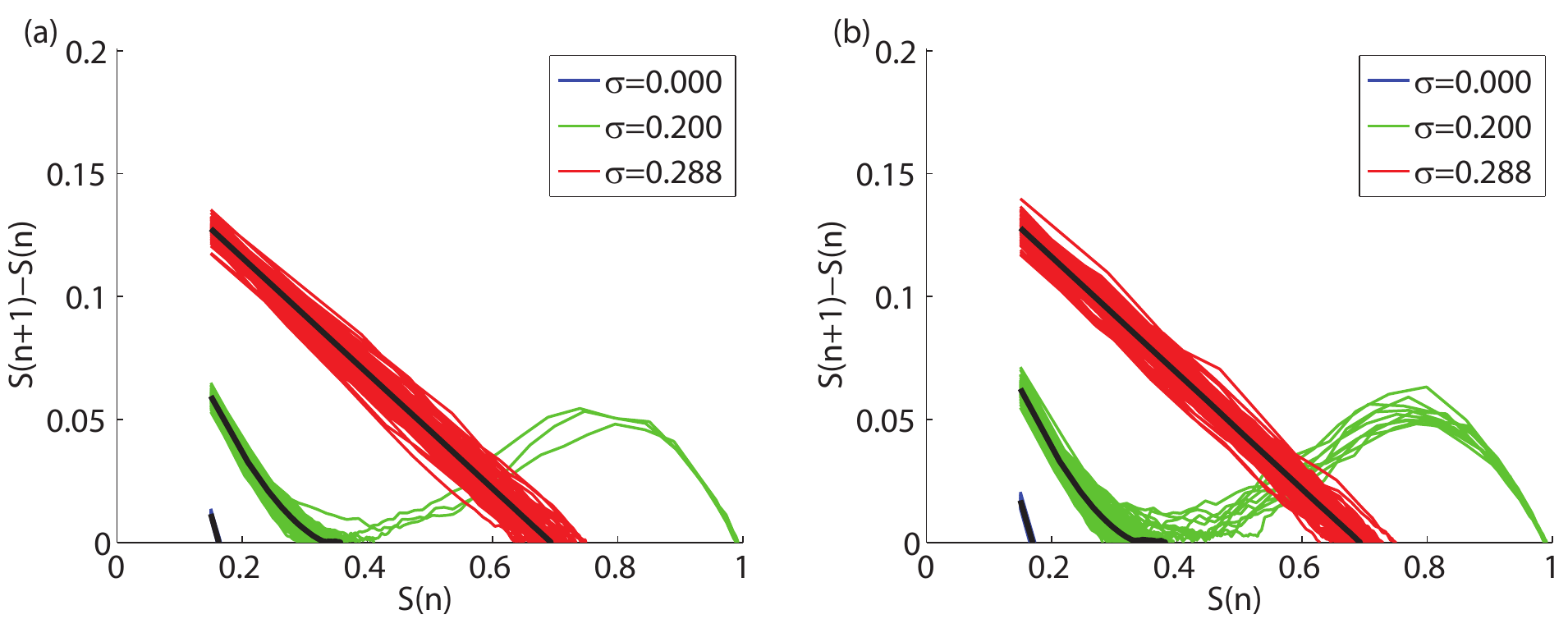}}
\vspace{0.3truecm}
\caption{{\bf Phase-space diagrams for a constant initiator fraction $p=0.15$},
and various standard deviations $\sigma=0$ (blue),
$\sigma=0.2$ (green), $\sigma=0.288$ (red) for (a) ER graphs and (b) SF networks with $\gamma=3$, with $z=10$ and $N=10^4$.
The colored lines refer to a hundred independent repetitions, while the black lines are their averages.}
\label{Fig7}
\end{figure}

\begin{figure}[h]
\centerline{\includegraphics[width=\textwidth]{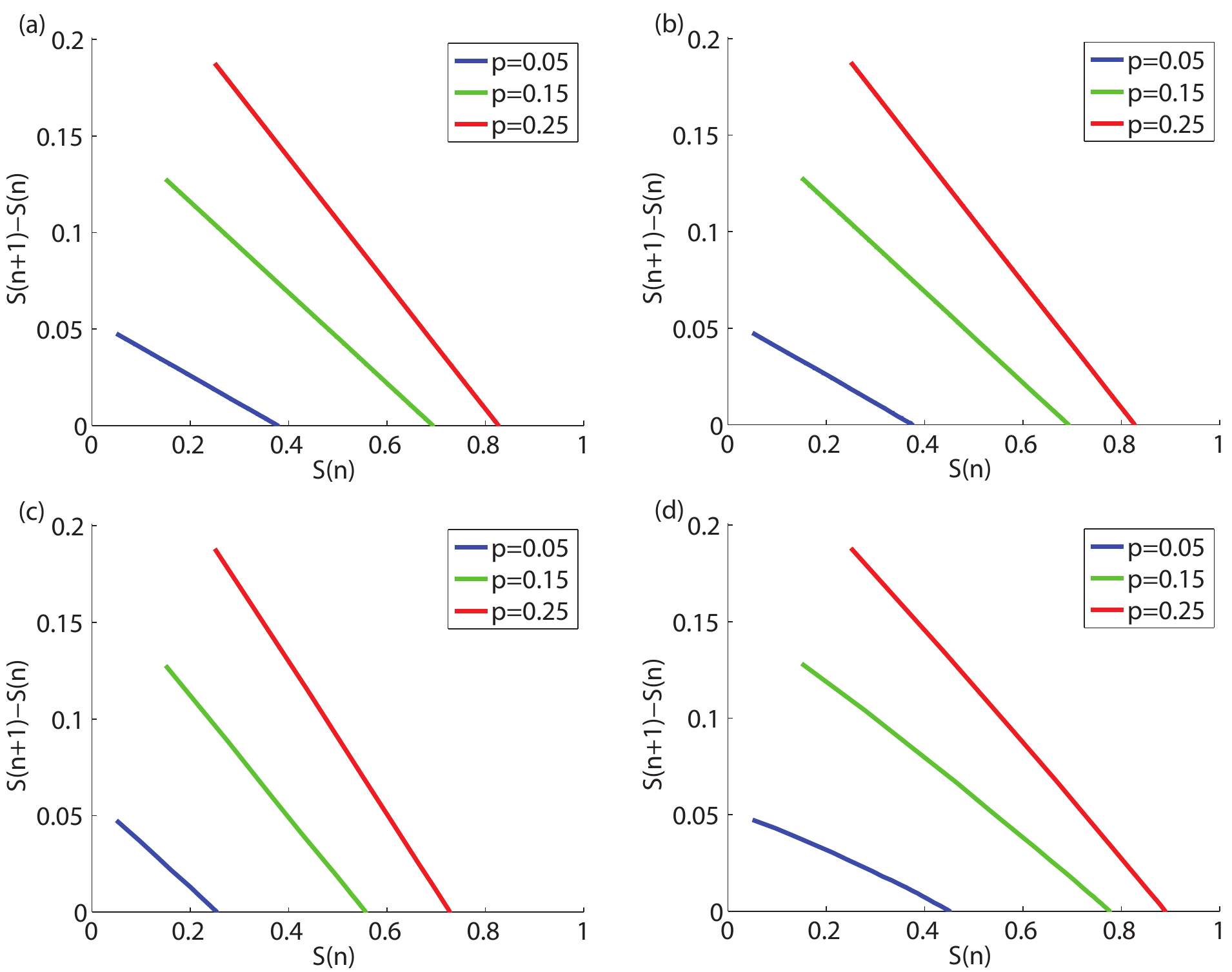}}
\vspace{0.3truecm}
\caption{{\bf Phase-space diagrams for the uniform random threshold distribution ($\sigma=0.288$)},
for various initiator fractions $p=0.05$ (blue), $p=0.15$ (red) and $p=0.25$ (green) for (a) ER graphs, (b) SF networks,
(c) high-school network, and (d) facebook network as in Fig.~\ref{Fig1}.
The solid lines and dotted lines (complete overlap) correspond to the simulations and to the closed-form analytic estimates [Eq.~(\ref{n_level_spread_TL})], respectively.}
\label{Fig8}
\end{figure}

To further understand the final cascade size behavior at the
critical point for synthetic graphs, we are examining the system
size dependence. The spread size at the equilibrium is independent
of the method of the insertion of initiators, e.g., it does not matter whether the addition
occurs in fractions or by individual addition of initiators. Using
Monte-Carlo simulations, Singh~\cite{Singh} showed that the  average
cascade size is largely independent of the system size for the same
initiator fraction for an identical threshold for ER graphs with unique
degree distribution. We use the same approach to show that this is
true for other threshold distributions for ER graphs (Fig.~\ref{Fig4})
and SF networks (Fig.~\ref{Fig5}). These results indicate that given an initiator fraction $p_0$ and an
average cascade size $S_{eq} \left(p_0\right)$, the addition of another
initiator fraction $p_1$ will cause the same change $\Delta
S=S_{eq}\left(p_0+p_1\right )-S_{eq}\left(p_0\right)$ in the average
cascade size $S_{eq}$, largely independently of the system size, for
large system sizes, for the same input degree and threshold
distributions.

\begin{figure}[h]
\centerline{\includegraphics[width=\textwidth]{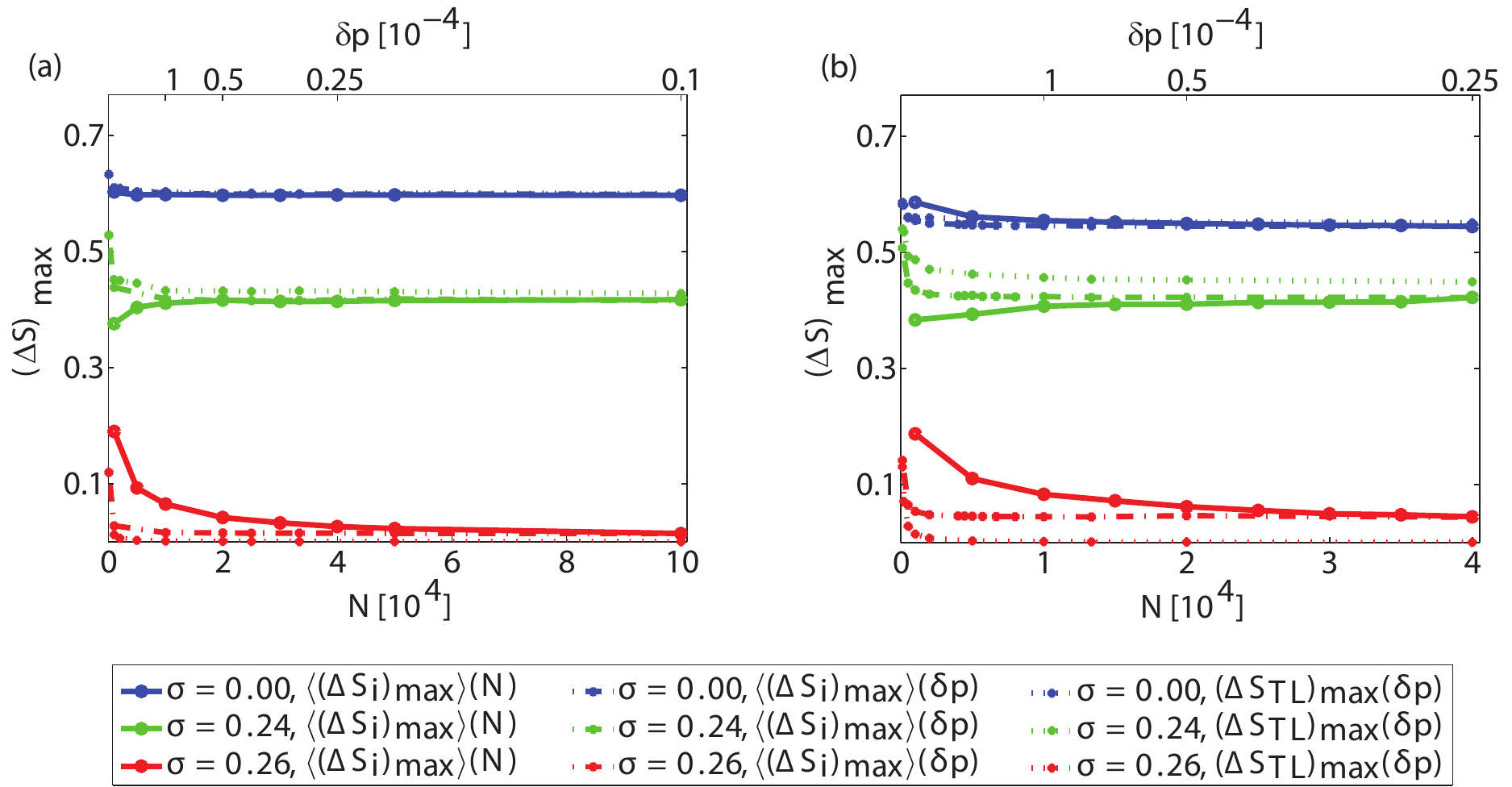}}
\vspace{0.3truecm}
\caption{{\bf Maximum contribution of initiators to the cascade size for various $\sigma$ values.}
(a) for ER graphs and (b) for SF networks with $\gamma=3$, for $z=10$.
Solid lines: $\langle (\Delta S_i)_{\text{max}}\rangle(N)$ of O(1) initiator with one-by-one addition of initiators for varying system sizes (bottom x-axis). Dashed lines: $\langle (\Delta S_i)_{\text{max}}\rangle(\delta p)$ for various initiator fractions (top x-axis) for a constant system size  $N=10^5$. Dotted lines: $(\Delta S_{TL})_{\text{max}}(\delta p)$ for various initiator fractions (top x-axis) for the TL approximation. The mean threshold is kept at $\phi_0=0.50$ in all cases.}
\label{Fig9}
\end{figure}

Our analysis so far focused on the cascade size at the steady state $S_{eq}$ averaged over many realizations of networks, threshold values and assignment of initiators (Figs.~\ref{Fig4} and \ref{Fig5}). To verify the presence and nature of phase transitions, we follow the approach presented in~\cite{Baxter_PRE2010}. We start by measuring the increase of the cascade size of each sample in response to the one-by-one addition of initiators. If a discontinuous phase transition arises, at the critical point, the increase of the cascade size should remain constant and independent of the system size. To investigate this, let $v$ be the current size of initiator set. For a given sample $i$, let $\Delta S_i=S_i (\frac{v+1}{N})-S_i (\frac{v}{N})$ denote the increase in the cascade size caused by the addition of a single randomly selected initiator to the current initiator set. Let $\left(\Delta S_{i}\right)_{\text{max}}\left(N\right)$ be the maximum value of $\Delta S_i \left(N\right)$  for all initiator sets of size $\frac{v}{N}$.  Then, varying $\sigma$, we study how $\left(\Delta S_{i}\right)_{\text{max}}\left(N\right)$ averaged over one thousand repetitions depends on the system size $N$ (Fig.~\ref{Fig9}) (solid lines).
We observe that for the plotted cases with $\sigma=0.00$ and $\sigma=0.24$, $\langle\left(\Delta S_i\right)_{\text{max}}\rangle\left(N\right)$ is independent of the system size. Moreover, the contribution of the rest of the initiators to the cascade tends to zero in the limit of infinite system size. However, for $\sigma=0.26$, $\langle\left(\Delta S_i\right)_{\text{max}}\rangle\left(N\right)$ decreases with the system size, indicating the absence of a discontinuous phase transition in the infinite system-size limit. Thus, there appears a qualitative change somewhere between $\sigma=0.24-0.26$.

A similar analysis can be applied to the analytical estimation, with the tree-like approximation, of the increase in the cascade size $(\Delta S_{TL})_{max}(\delta p)$ with a marginal addition of initiators. However, since the analytical estimation is set for an infinite system size, the one-by-one addition of initiators on larger and larger system sizes is not possible. Hence, we insert smaller and smaller fractions of initiators $\delta p$. In Fig.~\ref{Fig9} the top X axis is the fractional step increase of the number of initiators. For consistency, we include the corresponding increase in the cascade size $\langle\left(\Delta S_i\right)_{\text{max}}\rangle\left(\delta p\right)$ that $\delta p$, a fractional step increase of the number of initiators, measured through simulations. In this case, the minimum possible fraction of initiators is $\delta p=1/N$. We observe, that the results for the one-by-one addition of initiators with varying systems through simulations, agree with those for the fractional increase of an infinite system size with varying $\delta p$.  We conclude that it is between $\sigma=0.24-0.26$ (for $\phi_0=0.50$) where the discontinuous phase transitions cease to emerge in the thermodynamic limit.

\section*{Discussion}
Past experimental online studies~\cite{Latane_1996, Centola_2010, State_2015} indicate the existence of diverse
adoption thresholds of individuals in social networks. Prompted by
this observation, we studied the impact of diversity of thresholds
in spreading a new opinion, by intuitively assuming that the
adoption thresholds are drawn from a truncated normal distribution.
We explored this impact by using the threshold model, a
reinforcement model which has lately drawn significant attention in
the scientific community. We showed that in the presence of a
small spread (standard deviation) of the threshold distribution in a
network, unless a critical initiator fraction is reached, the impact
of the randomly selected initiators is small. Furthermore, we showed
that, when discontinuous transitions in
cascade size are possible for synthetic graphs, the addition of a single randomly-selected
initiator can have a significant (global) impact on the final
cascade size, i.e., the manifestation of the tipping point. However,
with a sufficiently large spread in the individual thresholds (with
the same mean), the cascade size exhibits a smooth transition, where
the impact of each added initiator is reduced by the current size of
the initiator set. Finally, we showed that in the case of a uniform
threshold distribution, the spreading rate is linearly decreasing
with the spread size for synthetic graphs and close to linearly
decreasing for empirical graphs. In summary, our results indicate
that information on the diversity of the thresholds is critically
important for the understanding of the behavior of cascades in
threshold-limited social contagion with multiple initiators. Most
importantly, sufficiently large spread in the individual thresholds
can change not only the quantitative aspects of triggering global
cascades, but also the qualitative behavior of the system: the
cascade size exhibits a smooth change (as opposed to a discontinuous
jump) as a function of the fraction of initiators.

\section*{Supporting Information}
\textbf{S1 text (pdf file)}
Includes basic statistics of the two empirical networks used and the closed-form analytical estimate for the case of the uniform distribution of thresholds.

\section*{Acknowledgments}
The authors are grateful to Ferenc Moln\'ar Jr. for his assistance on the generation of scale-free networks with the desired and accurate cutoffs and average degree \cite{Molnar_MDS}. 
Add Health was designed by J. Richard Udry, Peter S. Bearman, and
Kathleen Mullan Harris, and funded by a grant P01-HD31921
from the National Institute of Child Health and Human Development,
with cooperative funding from 17 other agencies. For data files
contact Add Health, Carolina Population Center, 123 W. Franklin
Street, Chapel Hill, NC27516-2524, addhealth@unc.edu.


\nolinenumbers

\section*{Author Contributions}
Conceived and designed the experiments: PDK SS BKS GK. Performed the experiments: PDK. Analyzed the data: PDK SS BKS GK.
Wrote the paper: PDK SS BKS GK.



%
%
%

\newpage

\newpage

\newpage

\newpage

\newpage

\newpage

\newpage

\newpage

\newpage

\newpage
\renewcommand{\thefigure}{S\arabic{figure}}\setcounter{figure}{0}
\renewcommand{\theequation}{S\arabic{equation}}\setcounter{equation}{0}

\renewcommand{\thesection}{S\arabic{section}}\setcounter{section}{0}


\title{The Impact of Heterogeneous Thresholds on Social Contagion with Multiple Initiators\\
{\bf Supporting Information}}
\author{PD Karampourniotis$^{1,2}\footnote{E-mail: karamp@rpi.edu}$, S Sreenivasan$^{1,2}$,  BK Szymanski$^{2,3}$, G Korniss$^{1,2}$}

\maketitle
\begin{flushleft}
$^{\bf{1}}$ Department of Physics, Applied Physics, and Astronomy, Rensselaer Polytechnic Institute,
110 8$^{th}$ Street, Troy, NY, 12180-3590 USA \\
$^{\bf{2}}$ Social Cognitive Networks Academic Research Center,
Rensselaer Polytechnic Institute, 110 8$^{th}$ Street, Troy, NY, 12180-3590 USA \\
$^{\bf{3}}$ Department of Computer Science,
Rensselaer Polytechnic Institute, 110 8$^{th}$ Street, Troy, NY, 12180-3590 USA \\
\end{flushleft}

\subsection*{Closed-form analytical estimate}

\label{S1_Text}
Here, we show explicitly the derivation of the closed form equation of the treelike approximation \cite{S_Gleeson_2007,S_Gleeson_time}
of the fraction $S_{n}$ of active nodes at level $n$ on  Eq.~(6) 
in the main text. According to \cite{S_Gleeson_time} the level (or time) dependent evolution of the fraction $q_{n+1}$ of nodes with inactive parents at level $n+1$ for synchronous updating of the nodes is given by
\begin{equation}
q_{n+1}=g(q_n)=p+\left(1-p\right)\sum\limits_{k=1}^\infty \frac{k}{z}P_k\sum\limits_{m=1}^{k-1} \binom {k-1} {m}q_n^m\left(1-q_n\right)^{k-m-1}F\left(\frac{m}{k}\right),
\label{S_q_n formula}
\end{equation}
and the fraction of active nodes at level $n+1$ is given by
\begin{equation}
S_{n+1}=h(q_n)=p+\left(1-p\right)\sum\limits_{k=1}^\infty P_k\sum\limits_{m=1}^k \binom {k} {m}q_n^m\left(1-q_n\right)^{k-m}F\left(\frac{m}{k}\right).
\label{S_{n+1} formula}
\end{equation}
The replacement of the cumulative probability function $F\left(\frac{m}{k}\right)$ in the
particular case of a uniform distribution of thresholds in the above two equations yields the closed form solution. Let a node $i$ have degree $k$ and an assigned threshold $\phi$. Vulnerability $l$ is the absolute number of active neigbhors required for node $i$ to get activated, and it is given by $l=\text{ceil}(\phi\times k)$. The cumulative probability distribution $F\left(\frac{m}{k}\right)$ of nodes with degree $k$, having vulnerability less or equal to $m$, is given by $F\left(\frac{m}{k}\right)=\sum\limits_{k=1}^m r_{l,k}$,  where $r_{l,k}$ is the probability that a node has vulnerability $l$, conditioned that it has degree $k$.
For a uniform threshold distribution the probability that a node has vulnerability $l$, conditioned that it has degree $k$, is $r_{(l,k)}=1/k$. For example, a node with degree $k=2$ will have vulnerability $l=1$, with probability $r_{\left(1,2\right)}=1/2$ and vulnerability $l=2$ with probability $r_{\left(2,2\right)}=1/2$. Thus, the fraction $F\left(\frac{m}{k}\right)$ of nodes that have vulnerability $m$ or less conditioned that they have degree $k$ for the uniform random threshold distribution is given by
\begin{equation}
F\left(\frac{m}{k}\right)=\sum\limits_{k=1}^m r_{l,k}=\sum\limits_{k=1}^m \frac{1}{k}=\frac{m}{k}.
\label{S_uni_F}
\end{equation}
Now, replacing Eq.~(S3) in Eq.~(S1) we show the linear relationship between the fraction $q_{n+1}$ of nodes with inactive parents at level $n+1$ with the fraction $q_n$ at the previous level $n$ of the approximated tree for
networks with uniform distribution of thresholds (see Eq~(3) in the
main text). So, 
\begin{equation}
q_{n+1} = p+\left(1-p\right)\sum\limits_{k=1}^\infty \frac{k}{z}P_k\sum\limits_{m=1}^{k-1} \binom {k-1} {m}q_n^m\left(1-q_n\right)^{k-1-m}\frac{m}{k},
\label{q_{n+1}:1}
\end{equation}
which simplifies to
\begin{equation}
q_{n+1} = p+\left(1-p\right)\frac{1}{z}\sum\limits_{k=1}^\infty P_k\sum\limits_{m=1}^{k-1} \binom {k-1} {m}q_n^m\left(1-q_n\right)^{k-1-m}m.
\label{q_{n+1}:2}
\end{equation}
However,
\begin{equation}
\sum\limits_{m=1}^{k} \binom {k} {m}q_n^m\left(1-q_n\right)^{k-m}m=\sum\limits_{m=0}^{k} \binom {k} {m}q_n^m\left(1-q_n\right)^{k-m}m,
\end{equation}
where the right hand of the equation is the mean of the binomial distribution, and it is given by $kq_n$~\cite{S_Grimmett}, thus
\begin{equation}
\sum\limits_{m=1}^{k-1} \binom {k-1} {m}q_n^m\left(1-q_n\right)^{k-1-m}m=\left(k-1\right)q_n
\end{equation}
Using the above equation in  Eq.~(S5) yields
\begin{equation}
q_{n+1} = p+\left(1-p\right)\frac{1}{z}\sum\limits_{k=1}^\infty P_k\left(k-1\right)q_n,
\label{q_{n+1}:3}
\end{equation}
which can be rewritten as
\begin{equation}
q_{n+1} = p+\left(1-p\right)\frac{1}{z}\left(\sum\limits_{k=0}^\infty P_k\left(k-1\right)+ P_0\right)q_n.
\label{q_{n+1}:5}
\end{equation}
Since the average degree is given by $z=\sum\limits_{k=0}^\infty kP_k$, the above equation becomes
\begin{equation}
q_{n+1} = p+\left(1-p\right)\frac{1}{z}\left(z-1+ P_0\right)q_n.
\label{q_{n+1}:6}
\end{equation}
which can be rewritten as
\begin{equation}
q_{n+1} = p+bq_n,
\label{q_n linear}
\end{equation}
with $b=\left(1-p\right)\frac{1}{z}\left(z-1+ P_0\right)$. The solution of the above equation with inititial condition $q_0=p$ is 
\begin{equation}
q_n = p\frac{1-b^{n+1}}{1-b}
\label{q_n sol}
\end{equation}

Similarly, replacing $F \left(\frac{m}{k} \right)$ in Eq.~(S2) formula 
by the right hand side of Eq.~(S3), the analytic approximation yields
\begin{equation}
S_{n+1} = p + \left(1-p\right)\sum\limits_{k=1}^\infty P_k\sum\limits_{m=1}^k \binom {k} {m}q_n^m\left(1-q_n\right)^{k-m}\frac{m}{k}.
\label{uni_final_cascade_size_TL:1}
\end{equation}
Using again the property of the mean of the binomial distribution  the above equation reduces to
\begin{equation}
S_{n+1} = p +\left(1-p\right)\sum\limits_{k=1}^\infty P_k\frac{1}{k}\left(kq_n\right),
\label{uni_final_cascade_size_TL:3}
\end{equation}
which yields
\begin{equation}
S_{n+1} = p +  \left(1-p\right)q_n\sum\limits_{k=1}^\infty P_k.
\label{S_{n+1}:4}
\end{equation}
Thus, the closed form solution of cascade size at level $n+1$ is given by
\begin{equation}
S_{n+1} = p + cq_n,
\label{S_{n+1} linear}
\end{equation}
with $c=\left(1-p\right)\left(1-P_0\right)$. 
Subtracting $S_n$ from both parts of the above equation and combining it with Eq.~(S11) we get
\begin{equation}
S_{n+1}-S_n = c\left(q_n-q_{n-1}\right).
\end{equation}

Substituting $q_{n} = p+bq_{n-1}$ from Eq.~(S11) into the above equation yields 
\begin{equation}
S_{n+1}-S_n = c\left(p+(b-1)q_{n-1}\right).
\end{equation}
Solving Eq.~(S16) for $q_{n-1}$ at level $n-1$ and substituting to the above equation yields 
\begin{equation}
S_{n+1}-S_n =c\left(p+(b-1) \left(\frac{S_n-p}{c}\right) \right).
\end{equation}
Expansion of the above equation yields to the closed form phase-space equation at Eq.~(6) 
in the main text
\begin{equation}
S_{n+1}-S_n =cp-(1-b) p-(1-b)S_n.
\end{equation}
Now, going back to the calculation of $S_{n+1}$ at Eq.~(S16), substituting $q_n$ with the right part of Eq.~(S12) yields
\begin{equation}
S_{n+1} = p + cp\frac{b^{n+1}-1}{b-1},
\end{equation}
where the cascade size $S_0$ at level $n=0$ is just the fraction of the initiators, $S_0=p$.
On the other hand, in the equilibrium state (as  $n\rightarrow \infty$) the cascade size $S_{eq}$ is given by
\begin{equation}
S_{eq} = p + cp\frac{1}{1-b},
\end{equation}
since $0\leq b<1$. Interestingly, the final cascade size doesn't depend for uncorrelated networks on the degree distribution, but only on the avearage degree z.


\end{document}